\documentclass[pre, reprint, superscriptaddress, amsmath, amssymb, floatfix]{revtex4-2}
\usepackage[dvipsnames]{xcolor}
\usepackage{float}
\usepackage{graphicx}
\usepackage{dcolumn}
\usepackage{bm}
 \usepackage{ulem}

\usepackage{cleveref}
\usepackage{xcolor}
\usepackage[protrusion=true,expansion=true]{microtype}
\usepackage{multirow}
\usepackage{makecell}

\begin{document}

\title{Optimal location of reinforced inertia to stabilize power grids}

\author{Sangjoon Park}
\affiliation{CCSS, KI for Grid Modernization, 
Korea Institute of Energy Technology, Naju, Jeonnam 58330, Korea} 

\author{Cook Hyun Kim}
\affiliation{CCSS, KI for Grid Modernization, 
Korea Institute of Energy Technology, Naju, Jeonnam 58330, Korea} 

\author{B. Kahng}
\affiliation{CCSS, KI for Grid Modernization, 
Korea Institute of Energy Technology, Naju, Jeonnam 58330, Korea}   
\date{\today}

\begin{abstract}
The increasing adoption of renewable energy sources has significantly reduced the inertia in the modernized power grid, making the system more vulnerable. One way to stabilize the grid is to add extra inertia from unused turbines, called the fast frequency response (FFR), to the existing grid.  
However, reinforcing inertia can cause unintended consequences, such as more significant avalanche failures. This phenomenon is known as the Braess paradox. Here, we propose a method to find the optimal position of FFR. This method is applied to the second-order Kuramoto model to find an effective position to mitigate cascading failures. To address this, we propose a method to evaluate a ratio between the positive effects of mitigation and the negative consequences. Through this analysis, we find that the peripheral area of the network is a seemingly effective location for inertia reinforcement across various reinforcement scales. This strategy provides essential insights for enhancing the stability of power grids in a time of widespread renewable energy usage.
\end{abstract}

\maketitle


\section{Introduction}
The power grid is essential for sustaining modern lives, including transportation, healthcare instruments, and food supply. Increasing environmental concerns have recently accelerated the adoption of renewable energy-based power generation systems~\cite{smith2022effect, dorfler2013synchronization, motter2013spontaneous,schmietendorf2017impact, song2022resilient, wei2019interdependence}. While this transition reduces dependence on fossil fuels, it has significantly decreased the inertia of power grids, leading to a desperate need for reinforcement strategies~\cite{nnoli2021spreading, rydin2020open}. Inertia plays a crucial role in reducing frequency fluctuations before control systems can respond, helping to prevent system instability. Without sufficient inertia, the grid becomes more vulnerable to frequency variations, resulting in large-scale blackouts~\cite{raman2022coupled, tayyebi2020frequency, fritzsch2024stabilizing}.

Various methods have been proposed to enhance the stability of power grids, such as increasing transmission line capacity or connecting new lines. However, these studies have shown that reinforcement may sometimes cause overloads, ultimately reducing grid stability rather than improving it~\cite{pagnier2019inertia, odor2024improving, schafer2022understanding, coletta2016linear, jhun2023prediction}. Similarly, while expected to enhance stability, inertia reinforcement may paradoxically exacerbate cascading failures, leading to unintended consequences.

This study investigates how the location and strength of inertia reinforcement affect cascading failure sizes in the swing equation, a sort of the second-order Kuramoto model~\cite{kuramoto1975international}. Our findings reveal that reinforcement can mitigate cascading failures but may also increase their failure size under certain situations. This underscores the importance of identifying optimal reinforcement locations. To evaluate the effects of reinforcement, we measure the ratio of mitigation to adverse outcomes. This ratio decreases as the severity of the Braess paradox increases; however, it increases when mitigation effects are substantial. 

By analyzing this ratio across various locations and reinforcement scales, we demonstrate that a topological metric, defined as the combination of mean shortest path length and degree, is a key indicator for identifying optimal reinforcement locations. The results indicate that this metric exhibits a high Pearson correlation coefficient~\cite{pearson1896vii} with the ratio, further supporting its relevance in reinforcement placement. To validate the effectiveness of this topological metric, we classify power plants into three groups based on their metric values: central group, peripheral group, and other group. We then analyze the distribution of the most effective reinforcement locations among these groups to determine which category contains the most optimal reinforcement sites.

The remainder of this paper is organized as follows: Section II introduces the second-order Kuramoto model and the avalanche process. The distribution of cascading failure sizes is obtained for various frequency thresholds. Section III defines inertia reinforcement strategies and examines their mitigation and adverse effects by measuring cascading failures. Section IV proposes a metric for evaluating reinforcement strategies and presents an optimized approach. Finally, Section V provides conclusions and discusses the implications of this study.

\section{Second-Order Kuramoto Model with Avalanche Process}
The generalized second-order Kuramoto model~\cite{kuramoto1975international} for phase and frequency synchronizations is expressed as follows: 
\begin{equation}
m_{i}\ddot{\theta}_{i}+\gamma_{i} \dot{\theta}_{i}= P_{i}+ \sum_{j\in {\rm n.n. of} i}^N{K_{ij}\sin(\theta_{j}-\theta_{i})},
\label{eq:swing}
\end{equation}
where $\theta_i$ and $\dot \theta_i$ represent the phase and angular velocity of bus $i$, respectively. The index $i$ denotes the bus index and runs $i=1,\cdots, N$. The parameters $m_i$ and $P_i$ denote the inertia of and power of bus $i$, respectively. $P_i$ can be positive for generators and negative for consumers. $\gamma_i$ is the dissipation coefficient, and $K_{ij}$ is the coupling strength between neighboring buses. As time passes, the system reaches a steady state where the power at each bus balances with the interaction term.

In the context of power grids, the second-order Kuramoto model is equivalent to the swing equation~\cite{odor2022synchronization, taher2019enhancing, odor2018heterogeneity}, where $K_{ij}$ depends on the specific buses $i$ and $j$ as $K_{ij} = V_i V_j Y_{ij}$. Here, $V_i$ and $V_j$ denote the voltages of buses $i$ and $j$, respectively, and $Y_{ij}$ is the admittance between them.

In real-world power grids, the frequency of each bus $\dot \theta_i$ can fluctuate due to mechanical faults, natural disasters, overload, etc. The frequency fluctuations can trigger cascading failures and large-scale blackouts~\cite{schafer2018dynamically, pahwa2014abruptness, daqing2014spatial, zhang2016optimizing}. We consider the avalanche process to account for the cascading failure: (i) Suppose a generator $i$ is malfunctioned. Then $P_i=0$. It can trigger frequency fluctuations in other buses. (ii) If a $\dot{\theta}_{j}$ exceeds a given threshold $\dot{\theta}_{\rm th}$, the bus $j$ is regarded unstable and set systematically $P_j\rightarrow 0$ for maintaining the remaining buses. This process repeats until no further overloaded bus remains. In this context, even if buses fail, the network topology remains unchanged during cascading failures because each bus in real power grid systems is equipped with circuits that separate the pathway connecting other buses and connecting the power generator.  We define the avalanche size $s_i$ as the number of shutdown buses during the cascading failure.

\begin{figure}
\centering
\includegraphics[width=0.99\linewidth]{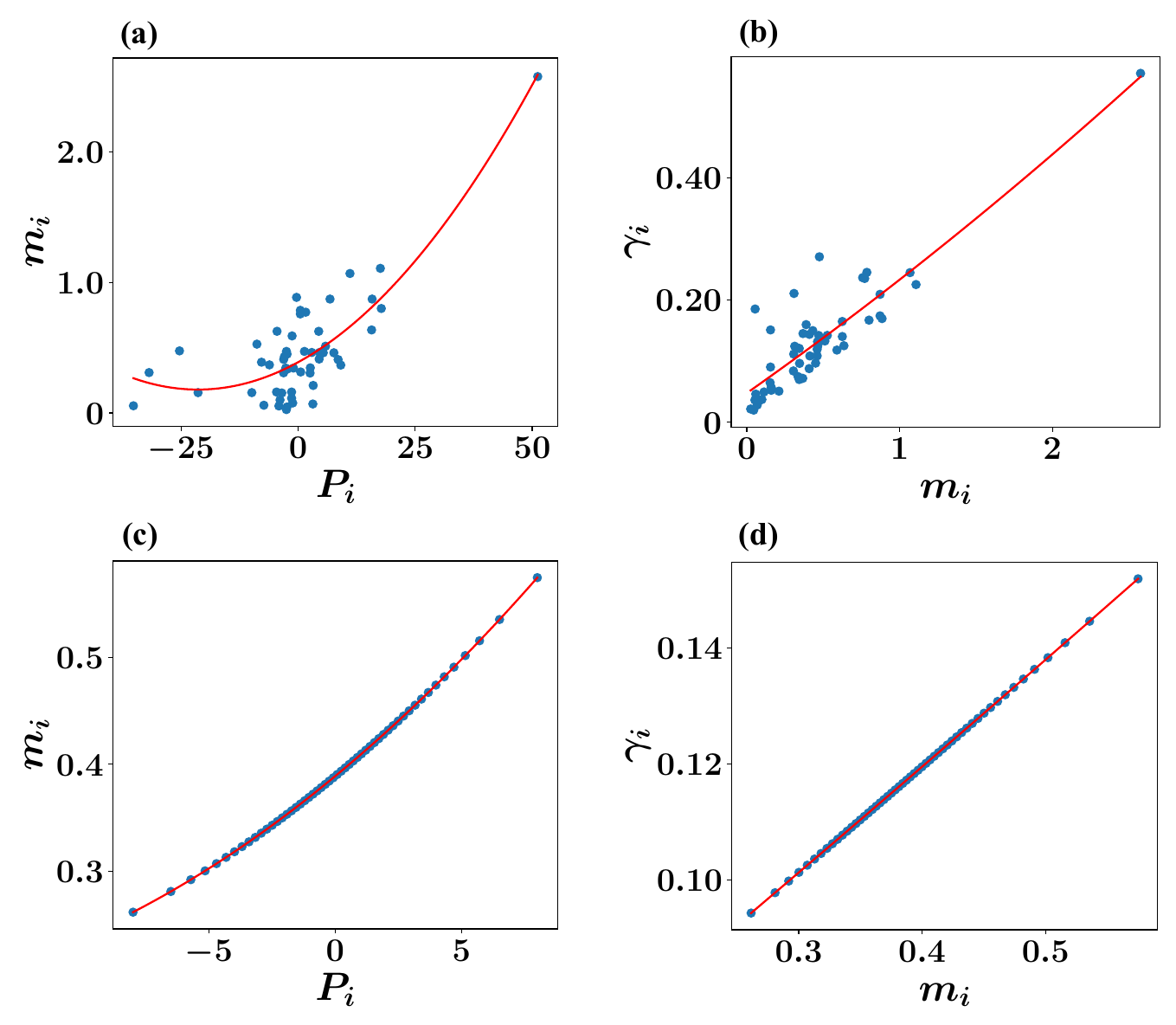}
\caption{Scatter plots illustrating the relationships between parameters in the real and modified UK power grids. (a) Inertia $m_i$ versus power $P_i$ for the real UK grid, with a quadratic fit (red curve) indicating that nodes with larger power generation tend to have higher inertia values. (b) Dissipation coefficient $\gamma_i$ versus inertia $m_i$ for the real UK grid, with a linear fit (red line) suggesting a proportional relationship between damping and inertia. (c) $m_i$ versus $P_i$ for the modified UK grid. (d) $\gamma_i$ versus $m_i$ for the modified UK grid. These parameter relationships are used to synthetically assign realistic values for $m_i$ and $\gamma_i$ based on the assigned $P_i$ values in our model.}
\label{fig1}
\end{figure}

The cascading dynamics depend on the topology of the network~\cite{yang2017small, rohden2016cascading, witthaut2015nonlocal, turalska2019cascading}. Thus, we adopted a real-world network, the UK power grid~\cite{lee2024reinforcement} which contains values for $m_i$, $\gamma_i$, $P_i$, and $K_{ij}$, for our studies. In this power grid, only a single configuration is available because all parameters are fixed. To explore various dynamics from different power configurations \{$P_i$\}, we synthetically set up a novel dataset for $P_i$ values sampled from a Gaussian distribution with zero mean and unit variance. The sampled values are adjusted to satisfy the constraint $\sum_i P_i = 0$ and subsequently normalized to lie within the interval [$-8$, $8$]. As shown in Fig.~\ref{fig1}(a) and (b), the parameters $m_i$, $\gamma_i$, and $P_i$ exhibit correlations. To capture these dependencies, we obtain fitting functions that describe the relationships between $P_i$, $m_i$, and $\gamma_i$. Based on these functions, we adjust $m_i$ and $\gamma_i$ to the sampled $P_i$ value (Fig~\ref{fig1}(c) and (d)).

\begin{figure}
\centering
\includegraphics[width=0.99\linewidth]{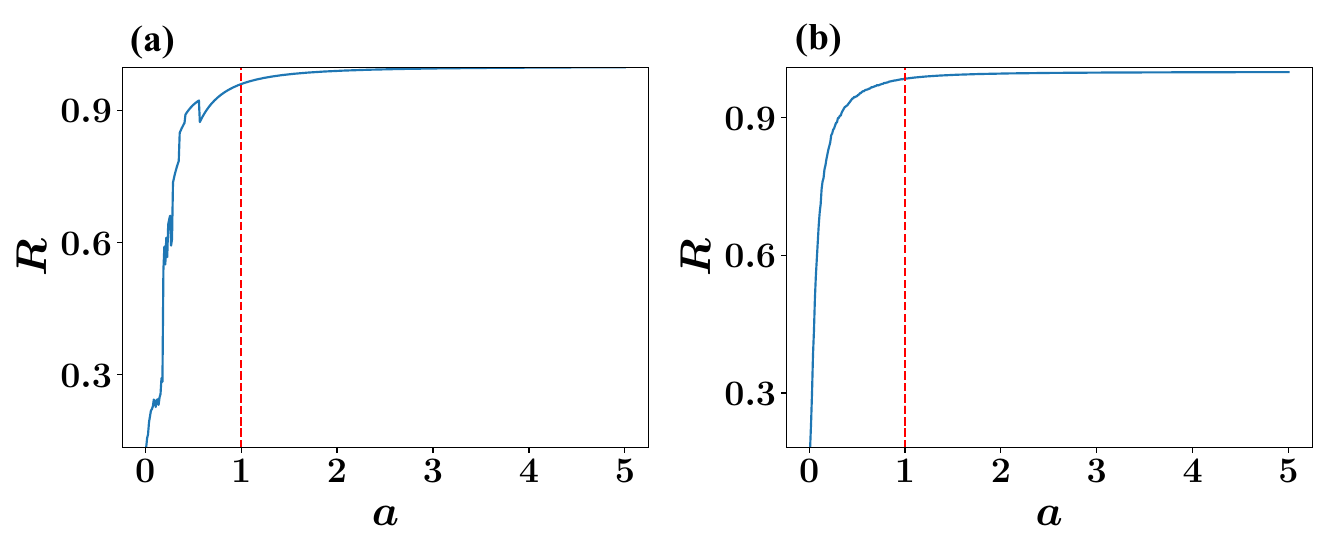}
\caption{Plot of the order parameter $R$ versus the scaling parameter $a$, which controls the coupling strength through the relation $K_{ij} = a V_i V_j Y_{ij}$. The red dashed vertical line indicates the specific value of $K$ (denoted as $K_R$) used in the main analysis, corresponding to the synchronized regime. (a) Results obtained using the real UK power grid dataset. (b) The powers $\{P_i\}$ are synthetically assigned from a Gaussian distribution, and the results are ensemble-averaged over 200 samples. The same UK grid topology is used for both cases. All simulations start from the initial condition $\theta_i = 0$ and $\dot{\theta}_i = 0$ for all oscillators. These results confirm that the system reaches a synchronized state under the parameter settings used in this study.}
\label{fig2}
\end{figure}

Prior to implementing the avalanche process, we check whether the system reaches a
synchronized state governed by Eq.~\eqref{eq:swing}. Using the parameter $a$ defined through $K_{ij}\equiv a V_i V_j Y_{ij}$, we control the coupling strength $K_{ij}$. The degree of synchronization among oscillators is quantified using the order parameter $R$ defined by $Re^{{\rm i}\Omega}=\sum_{i}e^{{\rm i}\theta_i}$, where $\Omega$ is the average angle. We check the behavior of the order parameter $R$ as a function of $a$. Fig.~\ref{fig2}(a) shows $R$ in the steady state as a function of $a$ using the UK real dataset. The vertical dashed line indicates the case $a=1$. Fig.~\ref{fig2}(b) shows $R$ vs. $a$ for the modified dataset, in which $\{P_i\}$ is replaced by the values chosen from the Gaussian distribution. The UK grid topology is used for (a) and (b). The vertical red dashed line indicates the case $a=1$. All simulations start from the same initial condition, where both $\theta_i = 0$ and $\dot{\theta}_i = 0$ for all oscillators. The red dashed line in Fig.~\ref{fig2} represents the $K$ used throughout the main analysis of this study, denoted as $K_R$. This value lies in the supercritical regime, indicating that the system under these parameter settings operates in a synchronized state.
In simulations, we consider the dynamics of the second-order Kuramoto model by using the Runge-Kutta 4th method with a time interval of $10^{-2}$ seconds.

\begin{figure}
\centering
\includegraphics[width=0.99\linewidth]{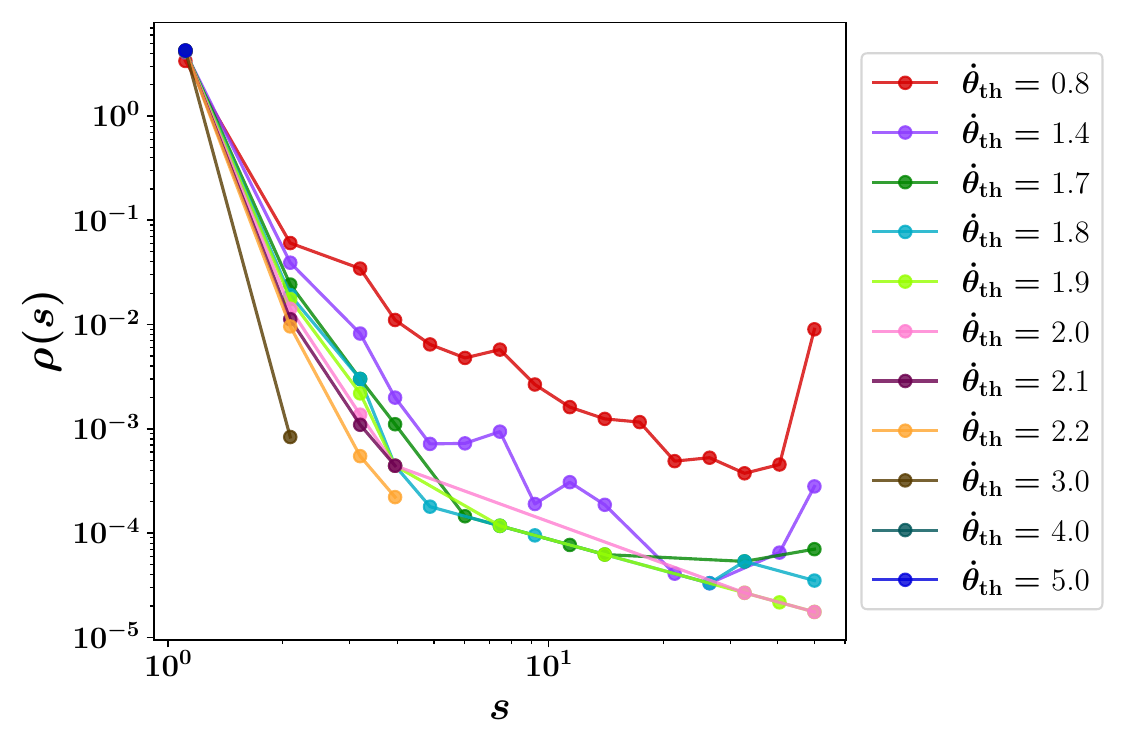}
\caption{Avalanche size distribution $\rho(s)$ as a function of avalanche size $s$ for different frequency thresholds $\dot{\theta}_{\rm th}$ in the second-order Kuramoto model. The simulations are implemented on the modified UK power grid with 200 different power configurations. The avalanche size distribution exhibits power-law behavior for large $s$ when the frequency threshold is taken as $\dot \theta_{\rm th}=1.9$.}
\label{fig3}
\end{figure}

The avalanche size distribution depends on the frequency threshold, $\dot{\theta}_{\rm th}$. As shown in Fig.~\ref{fig3}, the lower threshold generates a bump in the region of large avalanche size, indicating that the system is in a supercritical state. However, due to the limited system size ($N=54$), the subcritical behavior cannot be observed. We take $\dot{\theta}_{\rm th} = 1.9$, at which the avalanche size distribution exhibits seemingly diverse behavior~\cite{carreras2004evidence, carreras2016north, dobson2007complex, carreras2004complex}.

\section{Effects of Inertia Reinforcement and the Braess Paradox}
Many studies have shown that methods intended to improve stability may sometimes trigger the Braess paradox, where improvements lead to unintended adverse effects~\cite{pagnier2019inertia, odor2024improving, schafer2022understanding, coletta2016linear, jhun2023prediction}. This section deals with the case where inertia reinforcement can cause unintended adverse consequences. One standard method to reinforce inertia in real-world power grids is to use unused power plant turbines, known as synchronous condensers or fast frequency response (FFR). Specifically, suppose the inertia of generator $i$ ($P_{i} > 0$) with  $m_{i}$ is increased to $m_{\textrm {FFR}}$, where $m_{\textrm{FFR}}$ is variable, representing the magnitude of inertia reinforcement.  

\begin{figure*}
\centering
\includegraphics[width=0.99\linewidth]{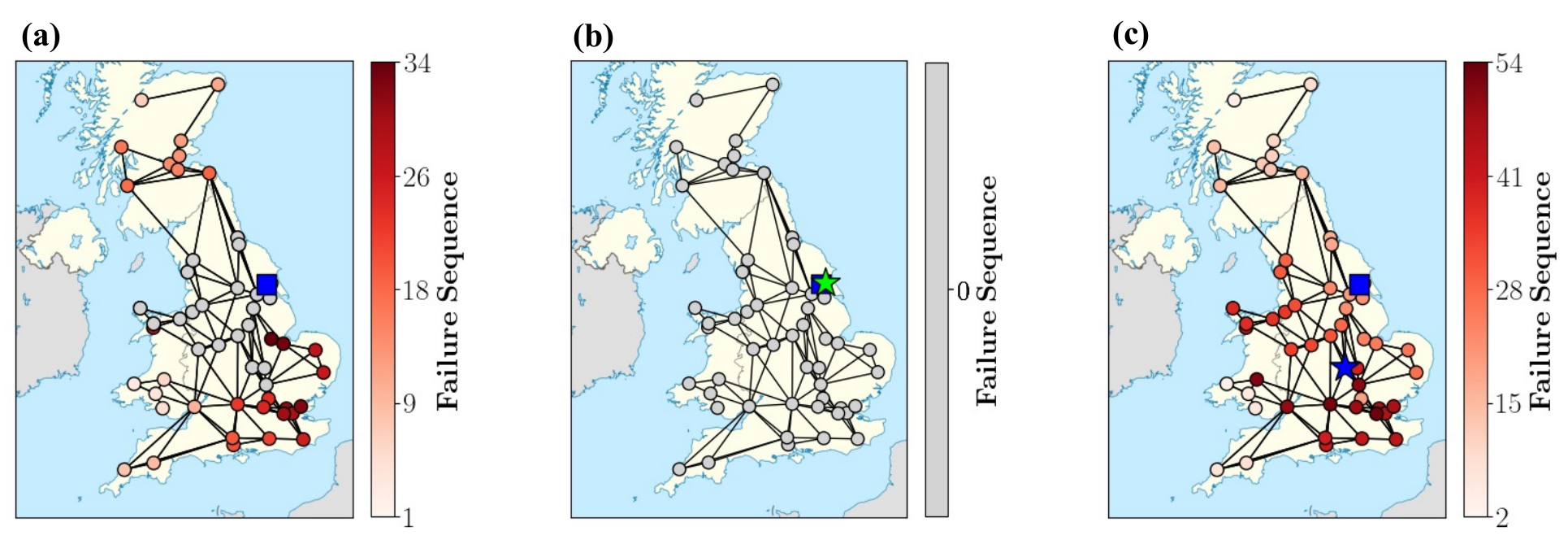}
\caption{When a node (\textcolor{blue}{$\blacksquare$}) is fault, frequencies of other nodes change. According to the threshold rule, nodes marked by \textcolor{red}{$\bullet$} subsequently fail; on the other hand, nodes marked by \textcolor{lightgray}{$\bullet$} remain survival. (a) shows the successively failed nodes when inertia reinforcement is absent. (b) shows that when the inertia reinforcement is implemented at (\textcolor{green}{$\bigstar$}), the avalanche size is reduced. (c) shows that when the inertia is reinforced to $m_{\rm FFR}=2$ at (\textcolor{blue}{$\bigstar$}), the avalanche size tremendously increases, indicating the Braess paradox.}  
\label{fig4}
\end{figure*}

We analyze the changes in cascading failure size based on the location of additional inertia when a fault occurs at the same power plant. Fig.~\ref{fig4} shows the cascading failure dynamics for $m_{\rm FFR} = 2$. A power plant fault triggers the cascading failures without reinforcing inertia (Fig.~\ref{fig4}(a)). Next, we examine whether the cascading failure size increases after inertia reinforcement compared to Fig.~\ref{fig4}(a). In Fig.~\ref{fig4}(b), reinforcement is applied to a node located in the east region. A comparison between Fig.~\ref{fig4}(a) and (b) reveals that the number of failed nodes decreases after reinforcement, indicating a mitigating effect. However, when reinforcement is applied at a different location, the cascading failure expands across the entire system, as shown in Fig.~\ref{fig4}(c). These results indicate that inertia reinforcement can sometimes lead to adverse effects contrary to general expectations. This highlights the importance of selecting the appropriate reinforcement location to avoid unintended consequences.

\section{Identifying Optimal Reinforcement Locations}
\subsection{Measure for the Effectiveness of Reinforcement}
We assess the extent of mitigation and adverse effects that arise at each potential reinforcement site to determine the optimal locations for inertia reinforcement. The cascading failure size, denoted as $s$, depends on the triggering power plant $i$ locations and the reinforced node $k$. The change in cascading failure size due to the reinforcement is defined as $\Delta s_{i}(k) \equiv s_{i}^{\rm after}(k) - s_{i}^{\rm before}$. When $\Delta s_{i}(k) < 0$, it indicates that the cascading failure size decreases after reinforcement, signifying successful mitigation. In contrast, $\Delta s_{i}(k) > 0$ suggests that inertia reinforcement has led to adverse effects, consistent with the Braess paradox. To evaluate reinforcement locations with minimal adverse effects and substantial mitigation, we calculate the ratio $r(k)$ for each node $k$ defined as follows:  
\begin{equation}
r(k)= {\sum_{i \in \Delta s_{i} < 0}{|\Delta s_{i}(k)|} \over {\sum_{j \in \Delta s_{j} > 0}{\Delta s_{j}(k)}}}.
\label{eq2}
\end{equation}
This ratio decreases as adverse effects become more significant or mitigation effects weaken. Nodes with higher $r(k)$ values are less vulnerable to the Braess paradox and more stable against cascading failures. Additionally, the reinforcement results are influenced by the power magnitude of the triggering power plant. To account for these variations, $r(k)$ is calculated over 200 ensembles of power configurations.

\subsection{Measure for the Optimal Reinforcement Location}

\begin{figure}[ht]
\centering
\includegraphics[width=0.99\linewidth]{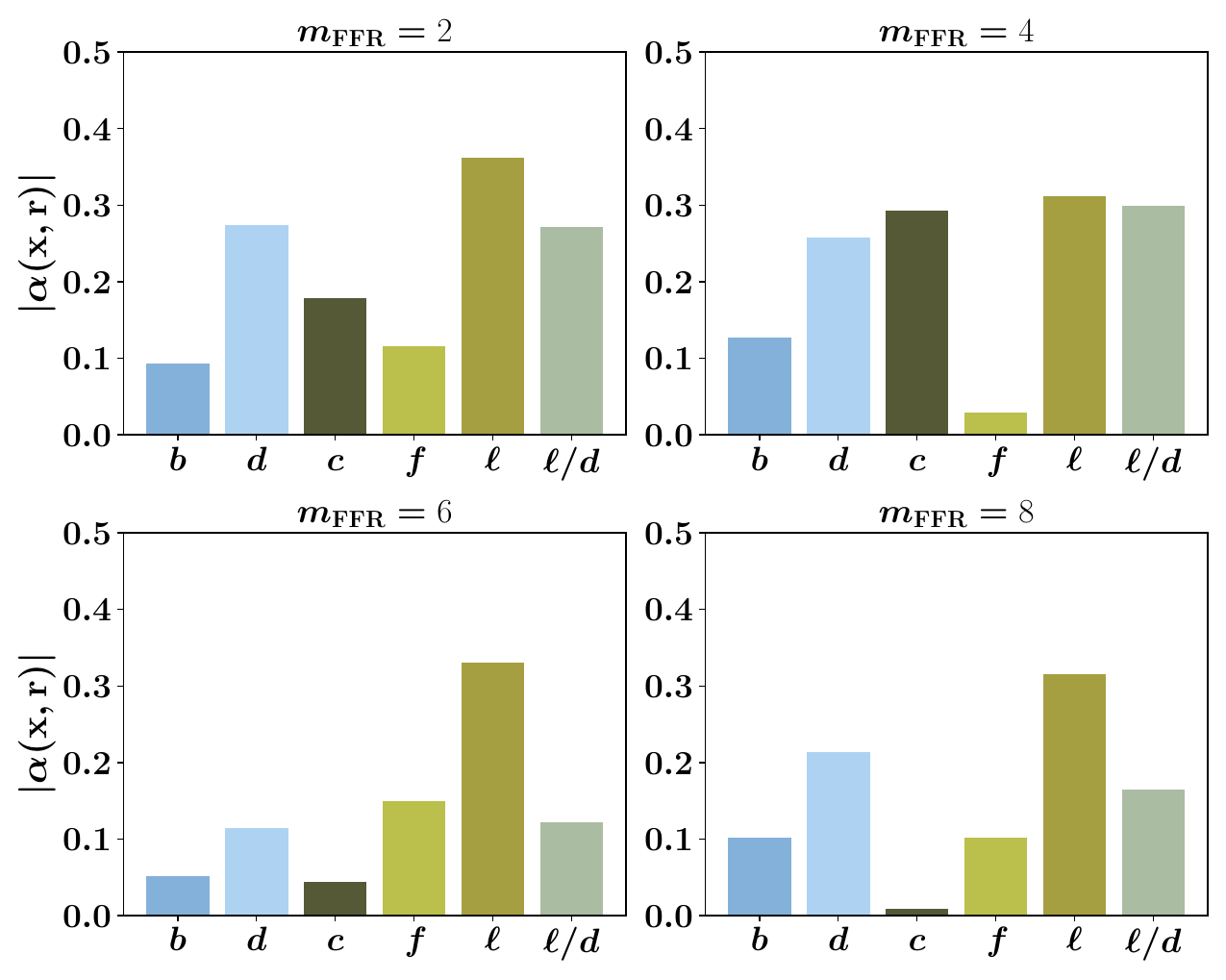}
\caption{The magnitude of the Pearson correlation coefficient ($\alpha$) between $\mathbf{r}$ and network metrics $\mathbf{x}$ of the reinforced node, including betweenness centrality ($b$), degree ($d$), clustering coefficient ($c$), eigenvector components of the Fiedler mode ($f$), mean shortest path length ($\ell$), and the combined metric ($\ell/d$). Panels (a), (b), (c), and (d) correspond to reinforcement sizes of $m_{\rm FFR} = 2, 4, 6, 8$, respectively. Each panel identifies the metric with the highest correlation, highlighting the consistent importance of the mean shortest path length ($\ell$) across all reinforcement scales.}
\label{fig5}
\end{figure}

\begin{figure}[ht]
\centering
\includegraphics[width=0.7\linewidth]{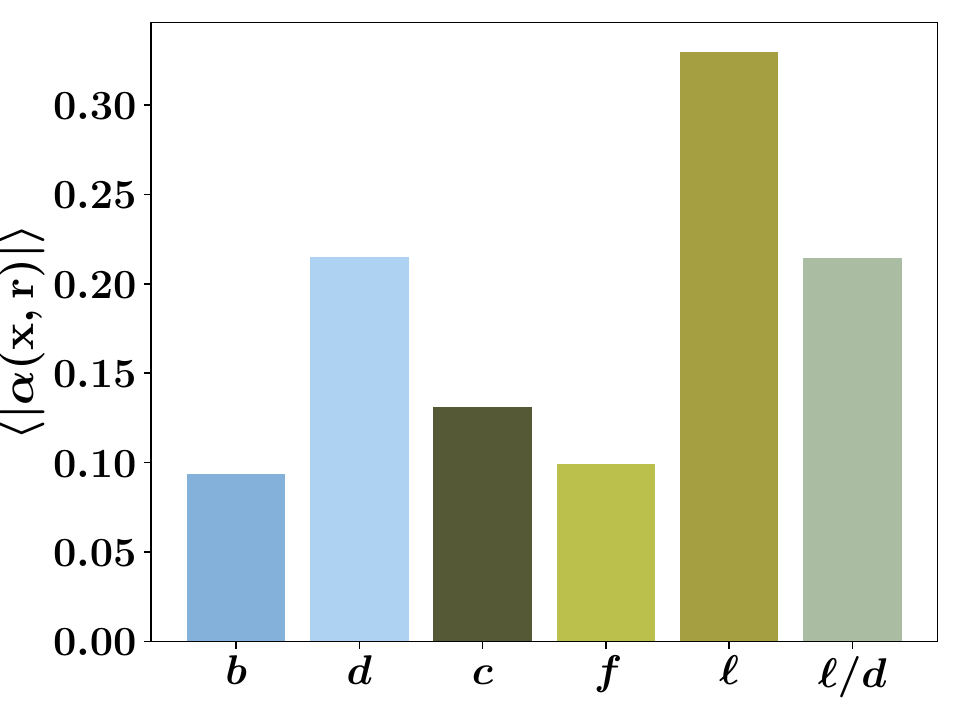}
\caption{Average magnitude of the Pearson correlation coefficient, denoted as $\langle |\alpha(\rm x, r)| \rangle$, between the reinforcement effectiveness ratio $\rm r$ and various network metrics $\rm x$. The considered metrics include betweenness centrality ($b$), degree ($d$), clustering coefficient ($c$), Fiedler component ($f$), mean shortest path length ($\ell$), and the combined metric ($\ell/d$). The results are averaged over various reinforcement magnitudes $m_{\mathrm{FFR}} = 2, 4, 6, 8$. Shows that $\ell$ exhibits the highest correlation with $\rm r$, indicating that it serves as the most reliable indicator for identifying effective reinforcement locations.}
\label{fig6}
\end{figure}

\begin{figure}[ht]
\centering
\includegraphics[width=0.8\linewidth]{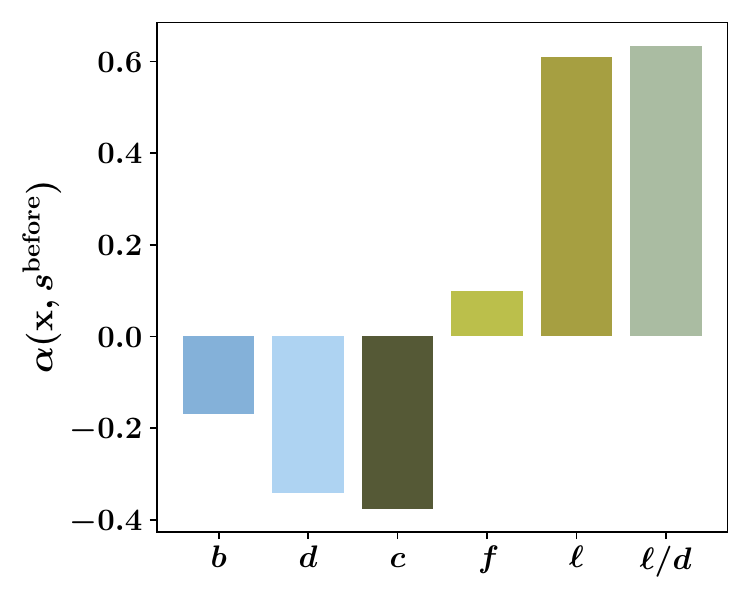}
\caption{The Pearson correlation coefficient ($\alpha$) between various network metrics ($\mathbf{x}$) and $s^{\rm before}$. Here, $\mathbf{x}$ represents different network metrics, including betweenness centrality ($b$), degree ($d$), clustering coefficient ($c$), eigenvector components of the Fiedler mode ($f$), mean shortest path length ($\ell$), and the combined metric ($\ell/d$). Among these metrics, $\ell/d$ and $\ell$ exhibit the high positive correlation, indicating that $\ell$ is strongly associated with avalanche sizes. This result underscores the significance of $\ell$ as a key metric for identifying vulnurable buses in the network.}
\label{fig7}
\end{figure}

\begin{figure*}[ht]
\centering
\includegraphics[width=0.99\linewidth]{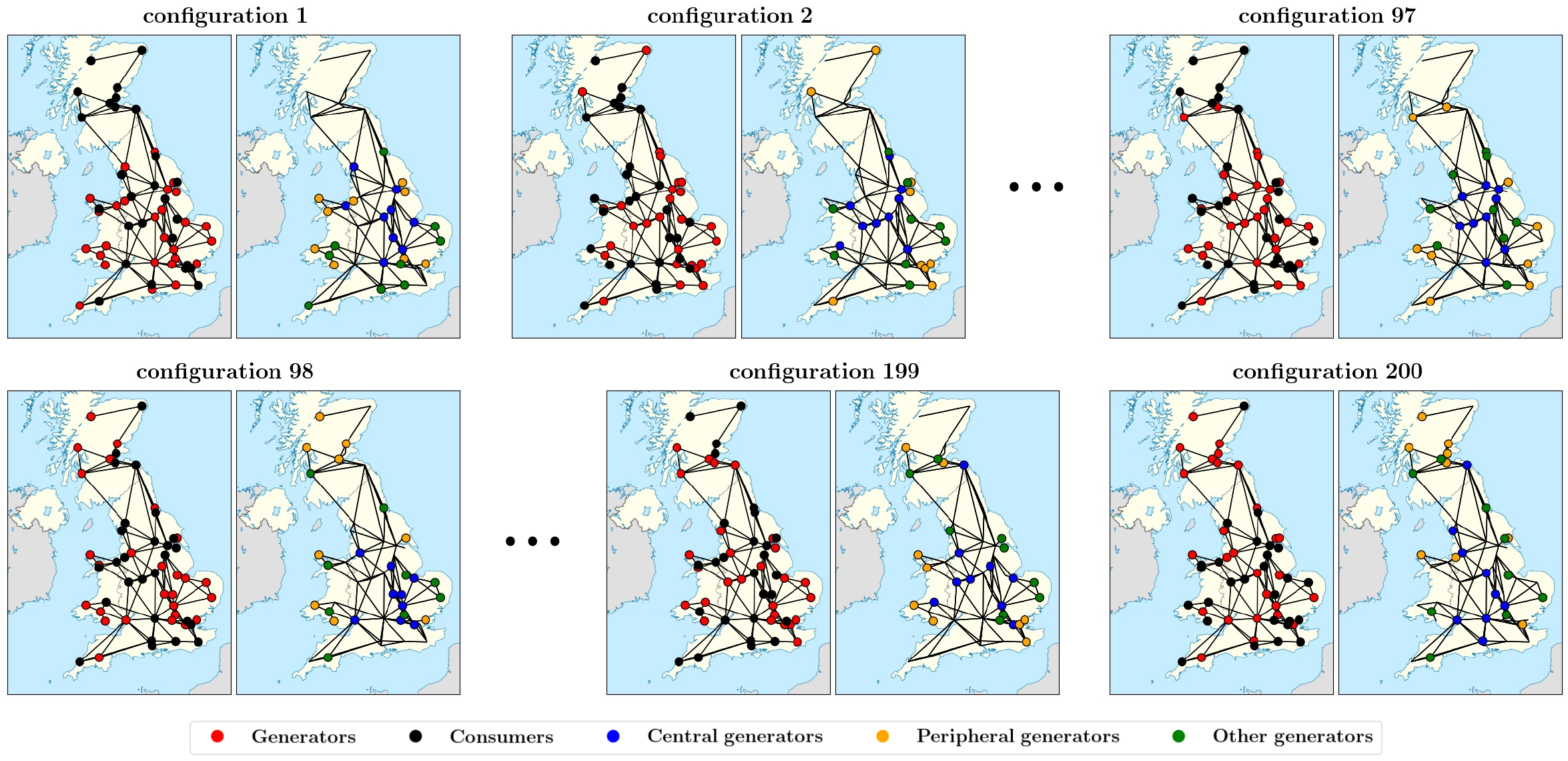}
\caption{Classification of power plants into three groups based on their $\ell_k$ values across the 200 ensembles of power configurations. The top 9 power plants with the highest $\ell_k$ values are categorized as the peripheral group (\textcolor{YellowOrange}{$\bullet$}), the bottom 9 as the central group (\textcolor{blue}{$\bullet$}), and the remaining power plants as the other group (\textcolor{Green}{$\bullet$}). Generators (\textcolor{red}{$\bullet$}) and consumers (\textcolor{black}{$\bullet$}) are also shown for reference.}
\label{fig8}
\end{figure*}

Considering an ensemble of power allocation configurations facilitates the analysis of how the topological properties of reinforcement locations suppress cascading failures. When analyzing a single configuration, it is difficult to determine whether the effectiveness of reinforcement is driven by the faulted plant's power magnitude or by the reinforcement site's topological effect. However, examining multiple power grid configurations reduces the impact of an initial power plant fault. For instance, if a reinforced power consistently leads to poor outcomes regardless of the fault power magnitude, the site is topologically weak. Therefore, we aim to identify key topological ingredients contributing to effective mitigation by analyzing reinforcement performance across various power grid configurations.

To investigate the topological properties of the most effective reinforcement locations, we measure the Pearson correlation coefficient $\alpha$ between $r(k)$ and various network metrics of the reinforcement node $k$. The considered metrics include betweenness centrality ($b$), degree ($d$), clustering coefficient ($c$), and the eigenvector components corresponding to the Fiedler mode ($f$). Additionally, we incorporate the mean shortest path length ($\ell$) across all nodes and the combined metric $\ell/d$, which accounts for network distance and connectivity. The Pearson correlation coefficient~\cite{pearson1896vii} is defined as:
\begin{equation}
\alpha(\mathbf{x}, \mathbf{r})=\frac{\frac{1}{N} \sum_{k=1}^N (r(k) - \overline{\mathbf{r}}) 
(x_k - \overline{\mathbf{x}})}
{\sqrt{\mathbb{V}(\mathbf{r}) \mathbb{V}(\mathbf{x})}},
\label{eq3}
\end{equation}
where $\mathbf{r} = [r(1), r(2), \dots, r(N)]$ and $\mathbf{x} = [x_1, x_2, \dots, x_N]$ represent the vectors of reinforcement effectiveness ratios and the network metrics, respectively. Here, $\mathbf{x}$ represents $b, d, c, f, \ell$, or $\ell/d$. For instance, if we use betweenness centrality as the network metric, $\mathbf{x}$ becomes $b = [b_1, b_2, \dots, b_N]$. $\mathbb{V}$ denotes the variance function, and $\overline{\mathbf{r}}$ and $\overline{\mathbf{x}}$ represent the mean values of the vectors $\mathbf{r}$ and $\mathbf{x}$, respectively.

The Pearson correlation coefficients for different reinforcement sizes ($m_{\rm FFR}=2,4,6,8$) are presented in Fig.~\ref{fig5}. The overall trend in Fig.~\ref{fig5} suggests that $\ell$ exhibits the strongest correlation. Furthermore, as shown in Fig.~\ref{fig6}, $\ell$ has the highest average magnitude of the Pearson correlation coefficient, $\langle |\alpha(\rm x, r)| \rangle$. These results indicate that $\ell$ is a key topological property and plays a crucial role in effective inertia reinforcement. To enable significant interactions between the reinforcement bus and other buses, the mean shortest path length should be small. Thus, the mean shortest path length $\ell$ should be small. The position with high $\ell$ is optimal for installment of FFR. This result is counterintuitive.
One may think that the bus with high degree and at the closer distance would be an optimal position; however, the plant at the position is robust due to the sharing load from neighbors plants, and the probability of the bus's failure is small. 
In Fig.~\ref{fig7}, we measure the Pearson correlation between $s^{\rm before}$ and various network metrics. The results indicate that $d$ negatively correlates with $s^{\rm before}$, and $\ell/d$ shows the highest positive correlation. This finding suggests that faults originating in nodes with lower degrees are more likely to trigger larger cascading failures. In other words, areas with many links tend to exhibit greater stability to disturbances than sparse regions. Consequently, reinforcement in robust regions often has minimal impact, as the avalanche size is naturally small, leading to lower $r(k)$ values. In contrast, vulnerable regions provide greater potential for improvement through reinforcement, resulting in higher mitigation effects and, consequently, higher $r(k)$ values.

\subsection{Classification of Effective Reinforcement Locations}

\begin{figure}[ht]
\centering
\includegraphics[width=0.99\linewidth]{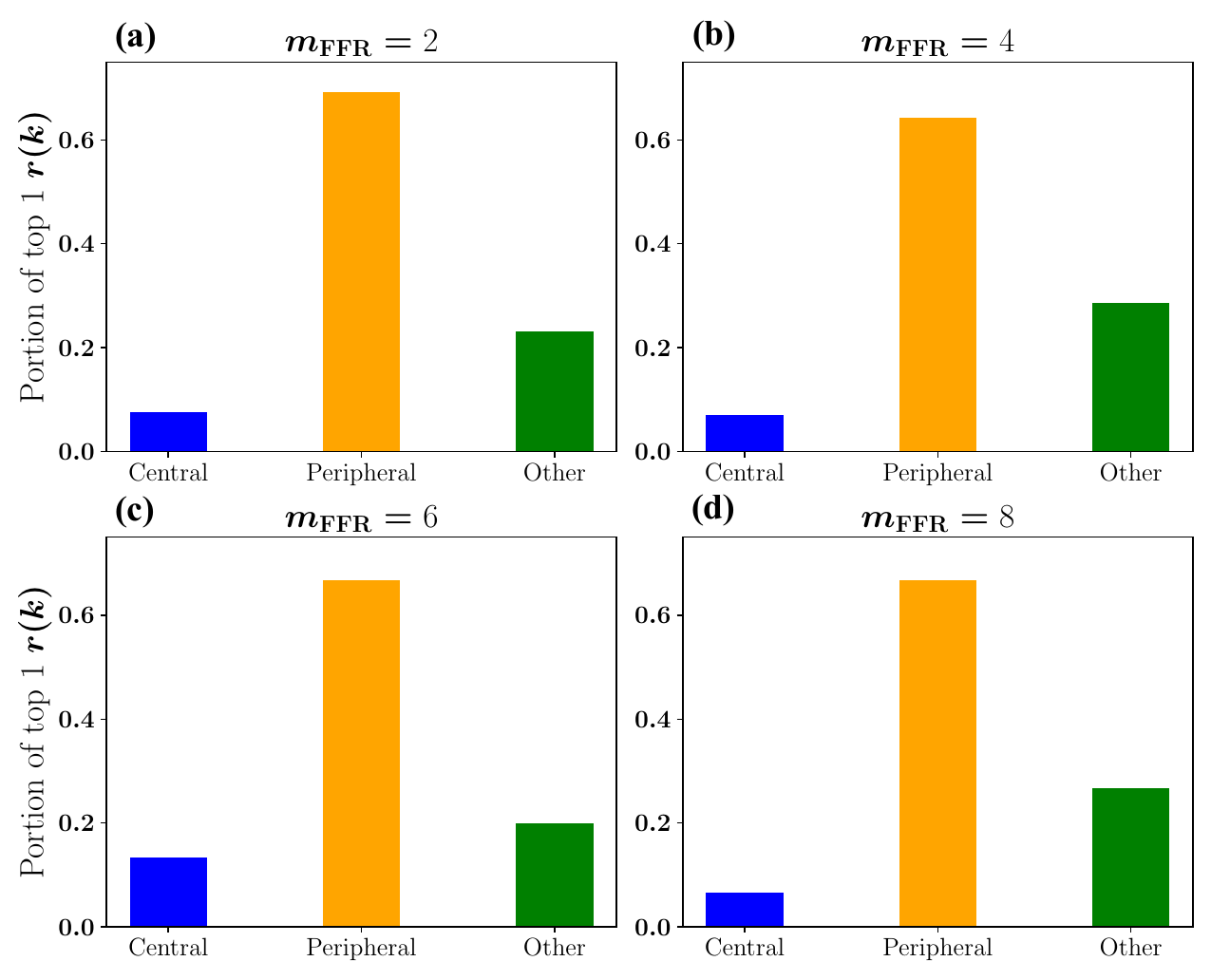}
\caption{Portion of power plants in each group (central, peripheral, and other) achieving the highest $r(k)$ across all configurations for different reinforcement sizes. Panels (a)–-(d) illustrate that the peripheral group nodes consistently dominate the top-performing locations for $m_{\rm FFR}=2,4,6,8$, emphasizing their effectiveness across all reinforcement scales.}
\label{fig9}
\end{figure}

To evaluate whether reinforcing nodes with high $\ell_k$ values leads to more effective mitigation, we classify power plants into three categories based on their $\ell_k$ values in each configuration. As shown in Fig.~\ref{fig8}, the top 9 power plants with the highest $\ell_k$ values are predominantly located near coastal regions; we refer to this category as the peripheral group. Conversely, the bottom 9 power plants, with the lowest $\ell_k$ values, are primarily located in relatively central regions and are designated as the central group. The remaining power plants, which fall between these two extremes, constitute the other category. To assess the effectiveness of this classification, we measure the portion of cases in which the most effective reinforcement locations belong to each group. As shown in Fig.~\ref{fig9}, across various reinforcement sizes ($m_{\rm FFR}=2,4,6,8$), the majority of reinforcement locations with high $r(k)$ values belong to the peripheral group. This result indicates that irrespective of the power allocation configuration, reinforcing power plants in the regions with high $\ell_k$ values consistently leads to lower adverse effects and greater mitigation efficiency.

\section{Conclusion}
We proposed a method to find the optimal position of the fast frequency response (FFR) installment to reinforce the inertia in the power grid. As previously remarked, we confirmed the adverse effect called the Braess paradox appears severely in power grids. Even though the inertia of a generator is increased, depending on the position, the cascading failure size can vary, even tremendously huge, leading to a blackout. Thus, finding the optimal position of the FFR is significant. Here, we proposed a system-wide method with the modified dataset from the real UK power grid, finding the optimal place of the FFR installation. 

We analyzed the ratio $r(k)$ to achieve this goal, reflecting mitigation effects and adverse outcomes. Our results show that nodes with high $r(k)$ values are predominantly in peripheral regions. This pattern remains consistent across different reinforcement scales $m_{\rm FFR}$. Peripheral nodes consistently demonstrate high mitigation effectiveness and minimal adverse effects, highlighting their importance in optimal reinforcement strategies.  

As renewable energy-based power plants proliferate, the need for effective inertia reinforcement will grow in future power grids. We anticipate that our proposed strategy will enhance the stability of next-generation grids with similar power generation characteristics.  

The second-order Kuramoto model effectively describes the frequency dynamics of power grids. Real-world power grids include threshold frequencies to protect equipment and disconnect nodes when significant frequency deviations occur. We incorporated the avalanche process into the second-order Kuramoto model to reflect this characteristic. Avalanche sizes are determined by the number of nodes exceeding the threshold frequency. As shown in Fig.~\ref{fig3}, the threshold frequency $\dot{\theta}_{\rm th}$ significantly influences avalanche dynamics, and we use the value where cascading failure sizes exhibit power-law behavior. 

In practical systems, synchronous condensers—idle power plant turbines—are often employed to reinforce inertia~\cite{nguyen2018combination, arayamparambil2020stabilising, nguyen2020applying}. We investigated the changes in cascading failures when inertia was reinforced using $m_k \rightarrow m_{\rm FFR}$. While inertia reinforcement is generally expected to improve stability, it can occasionally increase the scale of cascading failures, as observed in Fig.~\ref{fig4}. To address this, we calculated the ratio $r(k)$ for each reinforcement site $k$ to evaluate both mitigation and adverse effects. 

Given that power generation and consumption fluctuate continuously in real-world power systems~\cite{javaid2024energy, tang2015generation}, reinforcement locations determined under a specific power allocation configuration may become suboptimal as power conditions change. However, inertia reinforcement methods like synchronous condensers are not easily relocated in real-time. Therefore, an effective reinforcement strategy must ensure grid stability across varying power allocation configurations.

Since the topological properties of a power grid remain unchanged over time~\cite{komendantova2016social, ziaee2016optimal}, we propose a reinforcement strategy based on network topological metrics. As shown in Fig.~\ref{fig6}, $\ell$ exhibits the highest similarity with $r(k)$, establishing it as a reliable indicator for effective reinforcement. This result contrasts with the intuitive expectation that reinforcement should target nodes with a high degree and short distances to other nodes to maximize influence. In regions with high stability, failures are minimal, and thus, reinforcement is often unnecessary (Fig.~\ref{fig7}). Consequently, reinforcing locations that align with intuition results in low $r(k)$ values. This implies that reinforcing topologically vulnerable regions, rather than those with shorter average distances to other nodes, is critical for effective mitigation. To validate the effectiveness of reinforcement based on $\ell$, we analyze the distribution of the most effective reinforcement locations across different power configurations (Fig.~\ref{fig9}). The results reveal that, in most cases, the optimal reinforcement locations are found in the peripheral group, where $\ell$ values are highest. This finding underscores that reinforcing inertia in peripheral regions represents the most effective strategy for reducing the scale of cascading failures and enhancing overall grid stability.

During the preparation of this work the author(s) used ChatGPT o1 in order to language clarity and readability. After using this tool/service, the author(s) reviewed and edited the content as needed and take(s) full responsibility for the content of the publication.

\begin{acknowledgments}
B.K. was supported by the National Research Foundation of Korea by Grant No. RS-2023-00279802 and the KENTECH Research Grant No. KRG-2021-01-007.

\end{acknowledgments}

\providecommand{\noopsort}[1]{}\providecommand{\singleletter}[1]{#1}%

\end{document}